\documentclass[lettersize,journal]{IEEEtran}
\usepackage{amsmath,amssymb,amsfonts}
\usepackage{algorithmic}
\usepackage{array}
\usepackage[caption=false,font=normalsize,labelfont=sf,textfont=sf]{subfig}
\usepackage{textcomp}
\usepackage{stfloats}
\usepackage{url}
\usepackage{verbatim}
\usepackage{graphicx}
\usepackage{bm}
\usepackage{hyperref}

\usepackage[linesnumbered,ruled,vlined]{algorithm2e}

\graphicspath{{Figs/}}

\begin{document}
\title{Active RIS-Aided Anti-Jamming Wireless Communications: A Stackelberg Game Perspective}

\author{Xiao Tang, Zhen Ma, Bin Li, Cong Li, Qinghe Du, Dusit Niyato, and Zhu Han%
\thanks{X. Tang is with the School of Information and Communication Engineering, Xi'an Jiaotong University, Xi'an 710049, China, and also with Research \& Development Institute of Northwestern Polytechnical University in Shenzhen, Shenzhen 518063, China. (e-mail: tangxiao@xjtu.edu.cn)}
\thanks{Z. Ma and B. Li are with the School of Electronics and Information, Northwestern Polytechnical University, Xi'an 710072, China.}
\thanks{C. Li is with the National Key Laboratory of Science and Technology on Space Microwave, Xi’an 710100, China.}
\thanks{Q. Du is with the School of Information and Communication Engineering, Xi'an Jiaotong University, Xi'an 710049, China.}
\thanks{D. Niyato is with the College of Computing and Data Science, Nanyang Technological University, Singapore.}
\thanks{Z. Han is with the Department of Electrical and Computer Engineering, University of Houston, Houston 77004, USA, and also with the Department of Computer Science and Engineering, Kyung Hee University, Seoul 446-701, South Korea.}
}

\maketitle

\begin{abstract}
The pervasive threat of jamming attacks, particularly from adaptive jammers capable of optimizing their strategies, poses a significant challenge to the security and reliability of wireless communications. This paper addresses this issue by investigating anti-jamming communications empowered by an active reconfigurable intelligent surface. The strategic interaction between the legitimate system and the adaptive jammer is modeled as a Stackelberg game, where the legitimate user, acting as the leader, proactively designs its strategy while anticipating the jammer's optimal response. We prove the existence of the Stackelberg equilibrium and derive it using a backward induction method. Particularly, the jammer's optimal strategy is embedded into the leader's problem, resulting in a bi-level optimization that jointly considers legitimate transmit power, transmit/receive beamformers, and active reflection. We tackle this complex, non-convex problem by using a block coordinate descent framework, wherein subproblems are iteratively solved via convex relaxation and successive convex approximation techniques. Simulation results demonstrate the significant superiority of the proposed active RIS-assisted scheme in enhancing legitimate transmissions and degrading jamming effects compared to baseline schemes across various scenarios. These findings highlight the effectiveness of combining active RIS technology with a strategic game-theoretic framework for anti-jamming communications.
\end{abstract}



\section{Introduction}

The rapid development and proliferation of wireless communications enables various modern and advanced applications, for which the security and reliability of communications is of paramount importance~\cite{6g-sec}. However, the open and shared nature of the wireless medium makes wireless communications vulnerable to various attacks, among which, malicious jamming poses a significant threat by intentionally disrupting legitimate transmissions, downgrading service quality, and undermining communication integrity~\cite{anti-jam}. Moreover, as wireless devices become more sophisticated and intelligent, jammers become capable of adjusting their strategies by adapting to the wireless environment and even legitimate transmission behaviors. Such adaptive jamming attacks present an even more challenging security landscape, as jammers can dynamically learn and counteract existing defense mechanisms~\cite{intell-jam}. This escalating threat underscores an urgent and critical need for the development of intelligent and effective countermeasures to ensure the future integrity and reliability of wireless communications~\cite{intell-jam2}.

Meanwhile, reconfigurable intelligent surfaces (RIS) have emerged as a promising technology to revolutionize wireless communication by enabling controlled manipulation of the radio propagation environment~\cite{ris-sec}. Fundamentally, an RIS consists of a large array of low-cost and nearly-passive reflecting elements capable of inducing a controllable phase shift on incident electromagnetic waves~\cite{ris}. This allows for the creation of a smart radio environment, turning passive environmental elements into active contributors for enhancing communication performance~\cite{ris-jl}. Evolved from conventional passive RIS, active RIS have recently been proposed where the reflecting elements are also capable of signal amplification~\cite{a-ris}. The amplification capability is particularly important as it can compensate for the severe ``double path loss'' effect inherent in passive RIS systems, where the signal experiences both incident and reflection links~\cite{ris-jt}. This allows active RIS to not only redirect but also adjust incident signals, offering the potential to more actively intervene in wireless environments and communications~\cite{a-ris2}.

Due to its capability, an active RIS provides new opportunities to combat malicious jamming attacks for communication security provisioning~\cite{ris-jam1}. Rather, anti-jamming communications in conventional manners are usually achieved through resource diversity or signal processing, for which the required resource capacity and computational capability can be demanding to achieve effective mitigation~\cite{jam0}. In contrast, reflection-based transmission through RIS has changed the anti-jamming landscape by altering the propagation environment. However, conventional passive RISs may still suffer from legitimate power loss through double-fading attenuation and also inevitably bounce jamming signals to the receiver~\cite{ris-jd}. When further considering signal amplification through an active RIS, which not only reshapes the signal but also boosts the signal, enabling both power and spatial diversity to combat jamming attacks~\cite{a-ris-jam}. The active RIS may assist in overcoming the path loss of intended legitimate transmissions while downgrading malicious jamming signals, and thus allows a higher degree of freedom to achieve more effective anti-jamming communications~\cite{a-ris-jam2}.

Moreover, although anti-jamming communication is an extensively addressed issue in wireless research, existing work often relies on the assumption of static jamming behavior. This may deviate from the trend that wireless devices, including malicious ones, are becoming increasingly intelligent, which may sense and react to legitimate transmission behavior and improve the jamming effect~\cite{game-jam}. Towards this issue, we employ game theory to explicitly address the competitive situation between the legitimate and malicious sides, where game equilibrium analysis reveals the results from the interactions~\cite{game-jam2}. Recognizing the inherent sequential nature of interactive combating, the Stackelberg game, featuring a hierarchical structure, has become a natural and attractive model for anti-jamming communications~\cite{game-jam3}. When incorporating an active RIS into consideration, RIS reflection and amplification also need to interact with jamming behavior. This may significantly complicate anti-jamming game analysis due to the highly increased strategy space as compared to conventional models~\cite{game-jam4}.

Recognizing the potential of active RIS-assisted jamming mitigation, in this paper, we propose a Stackelberg game-based anti-jamming communication strategy while addressing the complicated interactions therein. Specifically, the main contributions of this work are summarized as follows:

\begin{itemize}
\item We consider active RIS-assisted communications in the presence of jamming attacks, for which we establish the Stackelberg game-based formulation with the legitimate and adversary sides being the leader and follower, respectively, with utility functions that incorporate the objective and cost of each side.
\item By characterizing the properties of the equilibrium, we develop a backward induction-based approach to derive the equilibrium. Then, for the jammer as the follower, we derive the optimal jamming policy for any given legitimate transmission and reflection behaviors.
\item With the optimal jamming policy incorporated as its reaction, the legitimate side jointly addresses transmit and receive beamforming, RIS reflection, and amplification to maximize the utility function. We decompose the joint optimization into subproblems within a block coordinate descent (BCD) framework with each decomposed subproblem solved and updated in an iterative manner to determine the anti-jamming strategy.
\item We present comprehensive simulation results to evaluate the performance of the proposed active RIS-assisted anti-jamming strategy. The results demonstrate significant improvements in the legitimate utility compared to benchmark schemes, which also validate the first-mover advantage gained from the Stackelberg formulation for security enhancement.
\end{itemize}

The rest of this paper is organized as follows. Sec.~\ref{sec:rw} reviews the related work. Sec.~\ref{sec:sys} explains the system model for the ARIS-aided communication scenario under jamming attack. Sec.~\ref{sec:game} presents the Stackelberg game formulation with equilibrium analysis. Sec.~\ref{sec:ris} derives the optimization for active RIS-assisted anti-jamming strategy design. Sec.~\ref{sec:sim} presents the simulation results, and Sec.~\ref{sec:con} concludes this paper.

\section{Related Work} \label{sec:rw}

\subsection{RIS-Assisted Anti-Jamming Transmissions}

The ability to manipulate the propagation environment of RIS empowers a new dimension for anti-jamming designs, and thus has attracted wide research interest. Initial research focuses on passive RIS, and the reflection is exploited with other resources for anti-jamming secure transmissions. In~\cite{ris-j4}, the authors deploy RIS for anti-jamming communications with a max-min fairness guarantee among users. In~\cite{ris-j2}, the authors exploit multiple RISs and jointly investigate RIS selection and beamforming for anti-jamming communications. In~\cite{ris-j1}, the authors propose an optimization-guided learning approach for RIS-assisted anti-jamming communications in an integrated terrestrial-satellite network. In~\cite{ris-j5}, the authors propose to use an unmanned aerial vehicle to mount an RIS to enable aerial jamming mitigation, with joint beamforming and location optimization. In~\cite{ris-j6}, the authors address RIS-based jamming attacks that break channel reciprocity for effective physical layer key generation. Besides conventional passive RISs, more recent research efforts have been devoted to emerging types of RIS for anti-jamming design. In~\cite{ris-j3}, the simultaneously transmitting and reflecting RIS is employed to combat both jamming and eavesdropping attacks. In~\cite{ris-j7}, the authors propose a stacked intelligent metasurfaces-assisted integrated-sensing-and-resistance anti-jamming scheme, which obtains jamming channel information in real time to assist the receiver in filtering out jamming. In~\cite{ris-j8}, the authors consider an intelligent omni-surface that simultaneously nullifies jamming and enhances desired signals by jointly manipulating its reflective and refractive properties to mitigate jamming attacks.

The reviewed literature manifests a growing interest in leveraging RIS technology for anti-jamming communications. While these contributions demonstrate the potential of RIS, many still concentrate on passive RIS, thereby not fully harnessing signal amplification to overcome significant path loss and strong jamming attacks. This motivates our research to investigate the role of active RIS, with its inherent amplification, for possibly more promising jamming mitigation.

\subsection{Game-Based Anti-Jamming Design}

Game theory provides a robust framework for modeling and analyzing strategic interactions in wireless communication systems, particularly in the presence of jamming-induced interest-conflicting players~\cite{HanGameBook}. In~\cite{game-j1}, the authors propose a high-altitude platform-assisted air-terrestrial networks framework to defend against jamming and eavesdropping; the coalition formation game is exploited for cooperative anti-jamming design. In~\cite{game-j5}, a jamming-combating power allocation strategy is proposed to secure data communication in IoT networks, which aims to minimize the worst-case jamming effect through a Colonel Blotto game-based model and analysis. More recently, RISs are also becoming more actively involved in game-based anti-jamming analysis and scheme design. In~\cite{game-j2}, the authors investigate an RIS-assisted satellite-terrestrial network faced with a smart jammer and propose a Stackelberg game-based anti-jamming design. In~\cite{game-j3}, the authors address the confrontation process between the communicator and the jammer as a Stackelberg game with RIS introduced as a helper; a slotted game analysis and jamming mitigation scheme is proposed based on equilibrium analysis. In~\cite{game-j4}, the authors investigate multi-layer RIS-transmitter-assisted anti-jamming communications from a Stackelberg game perspective; a learning-based approach is proposed to achieve the equilibrium to combat the attacks.

As seen, game theory is, by nature, effective in analyzing the strategic interactions inherent in anti-jamming scenarios. While recent studies have begun to integrate RIS into these game-based anti-jamming frameworks, a dedicated analysis of active RIS within such dynamic adversarial interactions remains less explored. Our work is therefore motivated by the need to develop a comprehensive Stackelberg game model that explicitly incorporates the amplification and reflection capabilities of active RIS, allowing the legitimate system to strategically leverage these advantages to combat an adaptive jammer.

\section{System Model} \label{sec:sys}

\begin{figure}[t]
    \centering
    \includegraphics[width=0.95\linewidth]{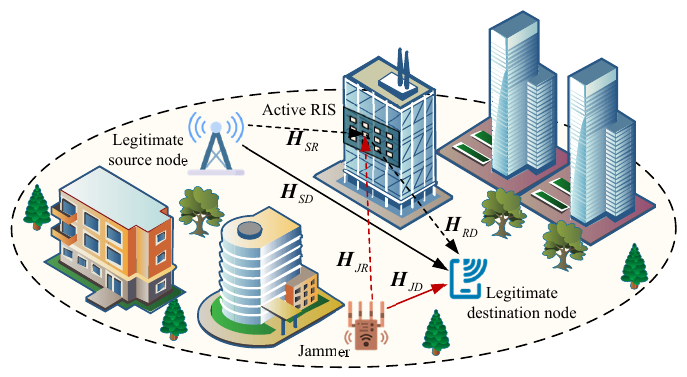}
    \caption{System model.}
    \label{fig:sys}
\end{figure}

We consider a wireless communication pair that consists a source node and a destination node, denoted by $ S $ and $ D $, respectively, where the legitimate transmissions are conducted in the presence of a malicious jammer, denoted by $J$. As the jammer intends to disrupt the legitimate communications by emitting a jamming signal, an active RIS, denoted by $R$, is deployed to enhance the legitimate transmissions while mitigating the jamming attacks by strategic amplification and reflection. The system is shown in Fig.~\ref{fig:sys}.

We assume the source, destination, and jamming nodes are equipped with $N_S$, $N_D$, and $N_J$ antennas, respectively. The Active RIS consists of $N$ reflecting elements, denoted by $\mathcal{N}=\{1,2,\ldots,N\}$,  each is capable of independent amplitude and phase adjustment of the incident signal. Let $\bm{H}_{SD} \in \mathbb{C}^{N_D \times N_S}$, $\bm{H}_{SR} \in \mathbb{C}^{N \times N_S}$, $\bm{H}_{RD} \in \mathbb{C}^{N_D \times N}$, $\bm{H}_{JD} \in \mathbb{C}^{N_D \times N_J}$, and $\bm{H}_{JR} \in \mathbb{C}^{N \times N_J}$ denote the channel matrices representing the direct link from $S$ to $D$, the link from $S$ to $R$, the link from $R$ to D, the jamming link from $J$ to $D$, and the link from $J$ to $R$, respectively. The active RIS employs a diagonal matrix $ \bm{\Theta} = \mathsf{\mathsf{diag}}(\bm{\theta}) $, where $ \bm{\theta} = \left[\alpha_n e^{j\vartheta_n}\right]_{n\in\mathcal{N}} $ with $ \alpha_n $ and $ \vartheta_n $ being the amplitude and phase shift of the $n$-the element, respectively. Also, the RIS amplification and phase shift are subject to the constraints that $ \alpha_n \le \alpha^{\max} $ and $ \vartheta_n \in [0, 2\pi] $, respectively, with $ \alpha_{\max} $ being the maximum amplification factor. Compared to its passive counterpart, an active RIS induces explicit power consumption and incurs higher hardware cost due to the integration of active amplifiers for each element. But allows additional degree of freedom and thus higher potential for anti-jamming designs.

For the considered system, the legitimate transmissions and jamming attacks are conducted simultaneously. Let $ x_s \in \mathbb{C}$ denote the transmit signal vector from $S$, and let ${x_J} \in \mathbb{C}$ denote the jamming signal vector from $J$, both are of unit power. The transmitter $S$ employs a transmit beamforming vector $\bm{w}_S \in \mathbb{C}^{N_S \times 1}$ with a power constraint $ \| \bm{w}_S \|^2 \le P_S^{\max} $, and the jammer $J$ uses a transmit beamforming vector $\bm{w}_J \in \mathbb{C}^{N_J \times 1}$ with a power limit $\| \bm{w}_J \|^2 \le P_J^{\max} $. Then, the received signal at the destination $D$, as a superposition of the signal transmitted from $S$ and the jamming signal from $J$, both via the direct and the reflected paths, along with the additive background Gaussian noise. As such, the received signal vector $\bm{y}_D \in \mathbb{C}^{N_D \times 1}$ is expressed as
\begin{equation}
\label{eq:yd}
\begin{split}
    \bm{y}_D = & \underbrace{(\bm{H}_{RD} \bm{\Theta} \bm{H}_{SR} + \bm{H}_{SD}) \bm{w}_S x_S}_{\text{Desired information signal}} \\
               & + \underbrace{(\bm{H}_{RD} \bm{\Theta} \bm{H}_{JR} + \bm{H}_{JD}) \bm{w}_J x_J}_{\text{Jamming signal}} \\
               & + \underbrace{\bm{H}_{RD} \bm{\Theta} \bm{n}_R}_{\text{Amplified RIS noise}} + \underbrace{\bm{n}_D}_{\text{Receiver noise}},
\end{split}
\end{equation}
where $\bm{n}_R \sim \mathcal{CN}(\bm{0}, \sigma_R^2 \bm{I}_N)$ represents the noise introduced by the active RIS elements, and $\bm{n}_D \sim \mathcal{CN}(\bm{0}, \sigma_D^2 \bm{I}_{N_D})$ is the additive white Gaussian noise at the receiver. To simplify the notation, we introduce the composite channel representations that incorporate the direct and reflection links as
\begin{align}
    \bm{H}_{SRD} \,\triangleq\, \bm{H}_{RD}\bm{\Theta}\bm{H}_{SR}, &\quad
    \bm{H}_{JRD} \,\triangleq\, \bm{H}_{RD}\bm{\Theta}\bm{H}_{JR}  \\
    \tilde{\bm{H}}_{SD} \,\triangleq\, \bm{H}_{SRD} \,+\, \bm{H}_{SD}, &\quad
    \tilde{\bm{H}}_{JD} \,\triangleq\, \bm{H}_{JRD} \,+\, \bm{H}_{JD}.
\end{align}
Here, we assume perfect channel state information is avaliable at both the legitimate side and the jammer. This follows a common modeling convention that faciliates a tractable game-theoretic analysis. In this regard, our model addresses the potential worst-case scenario from the leader's perspective, and the derived legitimate utility can be seen as a conservative performance bound.
Then, the achieved SINR for the legitimate transmissions under jamming attacks can be derived as
\begin{equation} \label{eq:sinr}
    \mathsf{SINR} = \frac{|\bm{w}_D^\dagger\tilde{\bm{H}}_{SD}\bm{w}_{S}|^2}
         {|\bm{w}_D^\dagger\tilde{\bm{H}}_{JD}\bm{w}_{J}|^2
         + \sigma_R^2\|\bm{w}_D^\dagger\bm{H}_{RD}\bm{\Theta}\|^2
         + \sigma_D^2\|\bm{w}_D^\dagger\|^2},
\end{equation}
where $ \bm{w}_D $ is the receive beamforming at the destination node with $ \left\|\bm{w}_D\right\|=1 $. For our considered anti-jamming model, the SINR serves as a fundamental performance metric for communications. Basically, SINR is intrinsically linked to achievable data rates (e.g., via the Shannon capacity formula) and communication reliability (e.g., Bit Error Rate). This inherent relationship makes SINR a natural and well-justified basis for constructing the utility functions that will govern the strategic decision-making of both the legitimate user and the jammer in the game-theoretic analysis detailed below.

\section{Game Formulation and Equilibrium} \label{sec:game}

To analyze the strategic interactions between the legitimate system and jammer, which aim for high-data-rate and reliable communications and attempt to disrupt the transmissions, respectively, we introduce the Stackelberg game formulation. The hierarchical game structure captures the sequential decision-making process between the adversaries, and the equilibrium reveals the results from their interactions.

\subsection{Stackelberg Game Formulation}

The Stackelberg game tracks the sequential actions between interest-conflicting players, including a leader which moves first and a follower that moves subsequently. For the proposed anti-jamming communication system, we consider the legitimate side takes the first action as the leader and the jammer acts as the follower. This setup follows the inherent nature of anti-jamming interactions, where the legitimate side determines the strategy first, establishing a signal that the jammer can then detect and launch jamming attacks. Thus, the jamming actions are fundamentally reactive, where the attack behavior is further optimized based on the properties of the ongoing legitimate transmissions.

Building upon the basic settings, we introduce the utility functions for the game players to establish a concrete game model. Basically, the utility functions are defined by considering the tradeoff between the desired outcome and cost for the players involved, aligning with their individual concerns. Specifically, for the jammer as the follower in the game, which intends to deteriorate the legitimate transmissions, the utility function is defined as
\begin{equation}
    u_J(\bm{w}_J; \bm{w}_S,\bm{w}_D,\bm{\Theta}) = -\mathsf{SINR}-c_J\| \bm{w}_J \|^2,
\end{equation}
where $ c_J $ is the coefficient of power cost. Here, we employ the negative of achieved SINR in~(\ref{eq:sinr}) as the reward for jamming, while taking the jamming power consumption into consideration. The utility function explicitly incorporates the legitimate-side strategy as an argument, as the achieved utility is affected by the interplay between the players. Accordingly, the jammer aims to maximize the achieved utility in the form of
\begin{IEEEeqnarray} {cl} \label{eq:prob-j}
    \IEEEyesnumber \IEEEyessubnumber*
    \max_{\bm{w}_J } & \quad  u_J(\bm{w}_J; \bm{w}_S,\bm{w}_D,\bm{\Theta}) \\
    \rm{s.t.} & \quad \| \bm{w}_J \|^2 \le P_J^{\max}.
\end{IEEEeqnarray}

Meanwhile, for the legitimate side, the primary goal is to achieve the highest-quality transmission, which can be effectively evaluated by the achieved SINR. Meanwhile, the legitimate transmission incurs a power consumption that needs to be considered as the cost. Accordingly, the legitimate-side utility function for the leader is defined as
\begin{equation}
    u_L(\bm{w}_S,\bm{w}_D,\bm{\Theta};\bm{w}_J) = \mathsf{SINR}-c_S\| \bm{w}_S \|^2,
\end{equation}
where $ C_S $ is the power cost coefficient, and we similarly incorporate the strategies of both sides as the arguments of utility function. Naturally, the legitimate side intends to maximize its own utility, while subject to the constraints imposed by the power budgets and reflection principles. Also, as the leader in the game, the legitimate side needs to incorporate the potential reaction of the follower into consideration to determine its own strategy. Therefore, the problem formulation for the active RIS-assisted legitimate system is expressed as
\begin{IEEEeqnarray} {cl} \label{eq:prob-l}
    \IEEEyesnumber \IEEEyessubnumber*
    \max_{\bm{w}_S,\bm{w}_D,\bm{\Theta}} & \quad  u_L(\bm{w}_S,\bm{w}_D,\bm{\Theta};\bm{w}_J) \\
    \rm{s.t.} & \quad \| \bm{w}_S \|^2 \le P_S^{\max},\quad \|\bm{w}_D\|^2 = 1, \label{eq:pwr-s-con}\\
    &\begin{aligned}\quad \|\bm{\Theta}\bm{H}_{SR}\bm{w}_S\|^2  &+ \|\bm{\Theta}\bm{H}_{JR}\bm{w}_J\|^2 \\& + \sigma_R^2\|\bm{\Theta}\|_F^2 \leq P_R^{\max}, \end{aligned} \label{eq:pwr-r-con}\\
    &\quad |\bm{\Theta}_{n,n}| \leq \alpha^{\max},\quad\forall n\in \mathcal{N}, \label{eq:refl-con}\\
    &\quad \bm{w}_J^\star = \arg\max_{\bm{w}_J}  u_J(\bm{w}_J; \bm{w}_S,\bm{w}_D,\bm{\Theta}), \label{eq:br-j-con}
\end{IEEEeqnarray}
where (\ref{eq:pwr-s-con}) and~(\ref{eq:pwr-r-con}) specify the power constraint at the source node and active RIS, respectively, and (\ref{eq:refl-con}) denotes the element-wise amplification constraint for the reflection, and~(\ref{eq:br-j-con}) indicates that the legitimate-side decision making needs to consider the jamming reaction. Note for an active RIS, the local thermal noise is also amplified and injected to the combined receive signal, thus the power budget of the RIS needs to jointly consider the legitimate signal, jamming attack, and noise, in order to enhance the desired signal against the self-induced noise amplification.

Based on the discussions above, we have formulated the Stackelberg game that the legitimate side and the jammer are associated with their own utility functions, which are individually maximized by optimizing their own strategies, as shown in~(\ref{eq:prob-j}) and~(\ref{eq:prob-l}). The sequential decision-making is defined as the legitimate side takes action first, followed by the reaction of the jammer. In accordance with the perfect information assumption, the jammer as the follower takes instant and optimal attack action. Meanwhile, the legitimate-side problem is defined by explicitly taking account of the jammer's potential reaction and the jammer simply takes the legitimate transmissions as a precondition to determine the jamming strategy. As a further note, our formulation can be conveniently extended to the scenario with imperfect information. In such cases, we explicitly introduce the channel uncertainty model and reformulate the problems in~(\ref{eq:prob-j}) and~(\ref{eq:prob-l}) with their robust counterparts, for which the robust strategies are required to solve the corresponding game.

\subsection{Equilibrium Analysis}

Since the players in the formulated Stackelberg game maximize their individual utility functions, the equilibrium is defined as the strategy profile that neither player will unilaterally deviate from if the other remains at its current strategy. Denote the equilibrium strategy profile as $ \left\{ \bm{w}_S^\star,\bm{w}_D^\star,\bm{\Theta}^\star, \bm{w}_J^\star \right\} $, and then it holds that
\begin{align}
    u_J(\bm{w}_J^\star; \bm{w}_S^\star,\bm{w}_D^\star,\bm{\Theta}^\star) \ge u_J(\bm{w}_J; \bm{w}_S^\star,\bm{w}_D^\star,\bm{\Theta}^\star),  \label{eq:eq-j}  \\
    \begin{aligned}
    u_L(\bm{w}_S^\star,\bm{w}_D^\star,\bm{\Theta}^\star, \bm{w}_J^\star(\bm{w}_S^\star,\bm{w}_D^\star,\bm{\Theta}^\star)) \qquad\qquad\:\:\\  \ge u_L(\bm{w}_S,\bm{w}_D,\bm{\Theta},\bm{w}_J^\star(\bm{w}_S,\bm{w}_D,\bm{\Theta})),  \label{eq:eq-l} \end{aligned} 
\end{align}
for any other strategy profile $ \left\{ \bm{w}_S,\bm{w}_D,\bm{\Theta}, \bm{w}_J \right\} $ that satisfies the constraints in~(\ref{eq:prob-j}) and~(\ref{eq:prob-l}). Here, the condition in~(\ref{eq:eq-j}) puts the jamming reaction as an explicit function of the legitimate strategy, which corresponds to the condition in~(\ref{eq:br-j-con}).

To derive the equilibrium, its existence must first be confirmed. For a one-leader-one-follower game, a Stackelberg equilibrium is guaranteed to exist if the follower's optimization problem yields a unique optimal solution for any given strategy of the leader \cite{E,E2}. We now demonstrate that this condition holds for our formulated game by analyzing the jammer's (follower's) problem.

We now analyze the jammer's optimization problem (\ref{eq:prob-j}) to determine its best response strategy $(\bm{w}_J^\star)$ for any given legitimate system strategy $(\bm{w}_S, \bm{w}_D, \bm{\Theta})$. To facilitate this analysis, we explicitly separate the jammer's transmit strategy into its power $P_J$ and its unit-norm beamforming direction vector $\hat{\bm{w}}_J$, such that the actual jamming beamforming vector is $\bm{w}_J = \sqrt{P_J}\hat{\bm{w}}_J$. The constraints are $0 \le P_J \le P_J^{\max}$ and $\|\hat{\bm{w}}_J\|^2 = 1$. Hereinafter, for notation simplicity while without ambiguity, we abuse the notation $\bm{w}_J$ to refer to the unit-norm direction vector when power $P_J$ is treated separately. Then, the jammer's problem in (\ref{eq:prob-j}) can then be recast as
\begin{IEEEeqnarray} {cl} \label{eq:prob-j-alt}
    \IEEEyesnumber \IEEEyessubnumber*
    \max_{P_J, \bm{w}_J } & \quad  u_J(P_J, \bm{w}_J) = -\mathsf{SINR}-c_J P_J \label{eq:prob-j-alt-obj} \\
    \rm{s.t.} & \quad 0\le P_J \le P_J^{\max}, \quad \| \bm{w}_J \|^2 = 1,
\end{IEEEeqnarray}
where the SINR is re-organized as
\begin{equation} 
    \mathsf{SINR} = \frac{|\bm{w}_D^\dagger\tilde{\bm{H}}_{SD}\bm{w}_{S}|^2}
         {P_J|\bm{w}_D^\dagger\tilde{\bm{H}}_{JD}\bm{w}_{J}|^2
         + \sigma_R^2\|\bm{w}_D^\dagger\bm{H}_{RD}\bm{\Theta}\|^2
         + \sigma_D^2}.
\end{equation}

For the problem in (\ref{eq:prob-j-alt}), we optimize $\bm{w}_J$ and $P_J$ sequentially. To maximize $u_J$, which means minimizing SINR, the jammer should choose its unit-norm beamforming direction $\bm{w}_J$ to maximize the term $|\bm{w}_D^\dagger\tilde{\bm{H}}_{JD}\bm{w}_{J}|^2$ in the denominator of the SINR. This is a maximum ratio transmission problem from the jammer's perspective, aligning its beam with the effective channel $\tilde{\bm{H}}_{JD}^\dagger\bm{w}_D$. Consequently, the optimal jamming beamforming direction is: 
\begin{equation} \label{eq:opt-j-bf}
    \bm{w}_J^\star = \frac{\tilde{\bm{H}}_{JD}^\dagger\bm{w}_D}
                         {\|\tilde{\bm{H}}_{JD}^\dagger\bm{w}_D\|},
\end{equation}
which is independent of the jamming power $P_J$. Next, substituting $\bm{w}_J^\star$ into $u_J(P_J, \bm{w}_J)$, we optimize $P_J$. The utility function in (\ref{eq:prob-j-alt-obj}) can be verified to be a concave function with respect to $P_J$. Therefore, the optimal $P_J$ can be found by setting the first-order derivative of $u_J$ with respect to $P_J$ to zero (if the unconstrained optimum is within $[0, P_J^{\max}]$) or by checking the boundary conditions. This yields the optimal jamming power: 
\begin{equation} \label{eq:opt-j-pwr}
    P_J^\star = \left[ \frac{\sqrt{P_S}|\bm{w}_D^\dagger\tilde{\bm{H}}_{SD}\bm{w}_{S}|}
        {\sqrt{C_J}\|\bm{w}_D^\dagger\tilde{\bm{H}}_{JD}\|}
    -\frac{\sigma_R^2\|\bm{w}_D^\dagger\bm{H}_{RD}\bm{\Theta}\|^2+ \sigma_D^2}
        {\|\bm{w}_D^\dagger\tilde{\bm{H}}_{JD}\|^2} \right]_0^{P_J^{\max}},
\end{equation}
where $(\cdot)_0^{P_J^{\max}} = \min\{\max\{0, \cdot\}, P_J^{\max}\}$ denotes projection onto the interval $[0, P_J^{\max}]$.

Based on the analysis above, we can see that the optimal jamming policy is uniquely determined for any given legitimate transmission strategy. Therefore, the existence of the Stackelberg equilibrium is confirmed. Meanwhile, the optimal jamming policy in~(\ref{eq:opt-j-bf}) and~(\ref{eq:opt-j-pwr}) provides the closed-form solution to the equilibrium condition in~(\ref{eq:eq-j}) at the follower's side. The legitimate-side equilibrium condition is further derived in the next section.

Since the optimal jamming strategy $\left\{\bm{w}_J^\star, P_J^\star\right\}$ is uniquely determined by (\ref{eq:opt-j-bf}) and (\ref{eq:opt-j-pwr}) for any given legitimate transmission strategy $\left\{\bm{w}_S, \bm{w}_D, \bm{\Theta}\right\}$, the condition for the existence of a Stackelberg equilibrium is satisfied. Also, these expressions for $\bm{w}_J^\star$ and $P_J^\star$ provide the closed-form solution for the follower's best response, satisfying the condition (\ref{eq:eq-j}). Then, the equilibrium analysis boils down to solve the leader's equilibrium condition in~(\ref{eq:eq-l}), as will be detailed in the subsequent section.

\section{Active RIS-Assisted Anti-Jamming Design} \label{sec:ris}

Generally, the Stackelberg game equilibrium is derived by employing the backward induction method. The follower's optimal reaction is first obtained, which is incorporated into the leader's problem to derive the leader's strategy. Then, the leader's optimal strategy is substituted into the follower's reaction to obtain the overall equilibrium. Here, the leader's consideration of the potential reaction of the follower reveals the sequential decision-making process of the game, where the leader's foresight supports the advantageous position of the first mover in the game.

\subsection{Reformulation and Decomposition}

The previous analysis has presented the optimal jamming policy, as the follower's reaction, on condition of given legitimate-side transmissions. We then employ the backward induction method to exploit the follower's strategy in the leader's problem to derive the leader's strategy. Specifically, for the leader's problem in~(\ref{eq:prob-l}), the follower's reaction in~(\ref{eq:br-j-con}) is now given in closed form as in~(\ref{eq:opt-j-bf}) and~(\ref{eq:opt-j-pwr}). To facilitate the analysis, we separate the legitimate beamformer as the transmit power $ P_S $ with unit-norm direction $ \bm{w}_S $ such that $ \quad P_S \le P_S^{\max} $ and $ \| \bm{w}_S \|^2 =1$. Hereinafter, the notation $ \bm{w}_S $ is abused to denote the legitimate beamforming direction. Accordingly, the leader's problem is reorganized as
\begin{IEEEeqnarray} {cl} \label{eq:prob-l-alt}
    \IEEEyesnumber \IEEEyessubnumber*
    \max_{P_S,\bm{w}_S,\bm{w}_D,\bm{\Theta}} 
    \quad& \begin{aligned} \frac{P_S|\bm{w}_D^\dagger\tilde{\bm{H}}_{SD}\bm{w}_{S}|^2}
                {P_J^\star|\bm{w}_D^\dagger\tilde{\bm{H}}_{JD}\bm{w}_{J}^\star|^2
                + \sigma_R^2\|\bm{w}_D^\dagger\bm{H}_{RD}\bm{\Theta}\|^2
                + \sigma_D^2} \\ -C_SP_S \end{aligned}  \nonumber\\
                \text{}\\
    \mathrm{s.t.}\quad& 0\le P_S\leq P_S^{\max}, \quad 0\le P_J\leq P_J^{\max}\\
    &\begin{aligned} P_S\|\bm{\Theta}\bm{H}_{SR}\bm{w}_S\|^2  &+ P_J^\star\|\bm{\Theta}\bm{H}_{JR}\bm{w}_J^\star\|^2 \\& + \sigma_R^2\|\bm{\Theta}\|_F^2 \leq P_R^{\max}, \end{aligned} \\
    &\|\bm{w}_D\|^2 = 1, \quad \|\bm{w}_S\|^2 = 1, \\
    &|\Theta_{n,n}| \leq \alpha^{\max}, \quad \forall n\in \mathcal{N} ,   \\
    &\{\bm{w}_J^\star,P_J^\star\} \text{~satisfies~}(\ref{eq:opt-j-bf})~(\ref{eq:opt-j-pwr}).
\end{IEEEeqnarray}

As can be seen, the leader's problem is rather complicated, particularly for constraints in the form of equality induced from the follower's best response. To facilitate the analysis, we explicitly substitute the follower's policy and obtain the reformulated problem as
\begin{IEEEeqnarray} {cl} \label{eq:prob-l-all}
    \IEEEyesnumber \IEEEyessubnumber*
    \max_{\substack{\bm{w}_S,\bm{w}_J,\bm{w}_D,\\P_S,P_J,\bm{\Theta}} }
    \quad&\frac{\sqrt{C_JP_S}|\bm{w}_D^\dagger\tilde{\bm{H}}_{SD}\bm{w}_{S}|}
    {\|\bm{w}_D^\dagger\tilde{\bm{H}}_{JD}\|}-C_SP_S  \label{eq:all-obj}\\
    \mathrm{s.t.}\quad&0\le P_S\leq P_S^{\max}, \quad 0\le P_J\leq P_J^{\max} , \label{eq:all-pwr-sj}  \\
    &\begin{aligned} P_S\|\bm{\Theta}\bm{H}_{SR}\bm{w}_S\|^2  &+ P_J\|\bm{\Theta}\bm{H}_{JR}\bm{w}_J\|^2 \\& + \sigma_R^2\|\bm{\Theta}\|_F^2 \leq P_R^{\max},\qquad \end{aligned} \label{eq:all-pwr-r} \\
    &\|\bm{w}_D\|^2 = 1,\quad \|\bm{w}_S\|^2 = 1 ,  \label{eq:all-beam-sd} \\
    &|\Theta_{n,n}| \leq \alpha^{\max},\quad \forall n\in \mathcal{N}  \label{eq:all-refl-amp} \\
   & \bm{w}_J^\star = \frac{\tilde{\bm{H}}_{JD}^\dagger\bm{w}_D}{\|\tilde{\bm{H}}_{JD}^\dagger\bm{w}_D\|}, \label{eq:all-beam-j} \\
     & \begin{aligned} P_J \geq  &\frac{\sqrt{P_S}|\bm{w}_D^\dagger\tilde{\bm{H}}_{SD}\bm{w}_{S}|}
        {\sqrt{C_J}\|\bm{w}_D^\dagger\tilde{\bm{H}}_{JD}\|} \\
    &-\frac{\sigma_R^2\|\bm{w}_D^\dagger\bm{H}_{RD}\bm{\Theta}\|^2+ \sigma_D^2}{\|\bm{w}_D^\dagger\tilde{\bm{H}}_{JD}\|^2} ,\end{aligned} \label{eq:all-pwr-j}
\end{IEEEeqnarray}
where the optimal jamming beamformer remains the same form in~(\ref{eq:all-beam-j}), and the newly introduced variable $ P_J $ induces tighter constraints in~(\ref{eq:all-pwr-r}) and~(\ref{eq:all-pwr-j}) as compared with their counterparts in~(\ref{eq:prob-l-alt}). As such, the problem in~(\ref{eq:prob-l-all}) achieves a lower-bound solution to the leader's original problem in~(\ref{eq:prob-l-alt}).

The reformulated leader's problem in~(\ref{eq:prob-l-all}) remains a challenging optimization task due to its non-convex nature and the intricate coupling among the optimization variables: the legitimate-side transmission, beamforming, and reflection, along with the jammer's optimal reactions. A direct joint optimization of all these variables is generally intractable. To address this complexity, we employ the BCD method, for which the fundamental motivation is to break down a complex, multi-variable optimization problem into a series of smaller, more manageable subproblems. Following this rationale, we decompose the leader's problem into three subproblems, each addressing a distinct aspect of the overall anti-jamming strategy: the first tackles the legitimate transmit and jamming powers, the second finds beamformers, and the third determines the RIS reflection. This decomposition allows us to handle the interdependencies among the variables, where the iterative updates to the solutions to the subproblems approximate the optimal legitimate strategy. The detailed solution methodologies developed for each of these subproblems are then elaborated in the subsequent subsections.

\subsection{Power Optimization}

The first subproblem within the BCD framework addresses the power optimization, for which we consider all other optimization variables, namely the beamforming vectors and the ARIS reflection coefficients to be fixed. Here, the jamming power is determined in a reactive manner, which is reflected by the substitution of jamming strategy into the legitimate side problem. Accordingly, the power optimization, as reduced from~(\ref{eq:prob-l-all}), is reduced as
\begin{IEEEeqnarray} {cl} \label{eq:subprob-power}
    \IEEEyesnumber \IEEEyessubnumber*
    \max_{P_S,P_J }
    \quad&\frac{\sqrt{C_JP_S}|\bm{w}_D^\dagger\tilde{\bm{H}}_{SD}\bm{w}_{S}|}
    {\|\bm{w}_D^\dagger\tilde{\bm{H}}_{JD}\|}-C_SP_S  \\
    \mathrm{s.t.} \quad& ~(\ref{eq:all-pwr-sj}),~(\ref{eq:all-pwr-r}),~(\ref{eq:all-pwr-j}). \nonumber
\end{IEEEeqnarray}
As we can readily observe from the objective function in~(\ref{eq:subprob-power}), a non-convexity is introduced by the square root operation on the legitimate transmit power. To address this issue and transform the problem into a more tractable form, we introduce a new optimization variable $ \gamma=\sqrt{P_S} $. Consequently, the transmit power is expressed as $ P_S = \gamma^2 $. Then, substituting $P_S = \gamma^2$ into the problem, it can be rewritten in terms of $\gamma$ and $P_J$ as follows
\begin{IEEEeqnarray} {cl} \label{eq:subprob-pwr}
    \IEEEyesnumber \IEEEyessubnumber*
    \max_{\gamma,P_J }
    \quad&\frac{\sqrt{C_J}\gamma|\bm{w}_D^\dagger\tilde{\bm{H}}_{SD}\bm{w}_{S}|}
    {\|\bm{w}_D^\dagger\tilde{\bm{H}}_{JD}\|}-C_S\gamma^2  \\
    \mathrm{s.t.} \quad& ~(\ref{eq:all-pwr-sj}), \nonumber\\
    &\begin{aligned} \gamma^2\|\bm{\Theta}\bm{H}_{SR}\bm{w}_S\|^2  &+ P_J\|\bm{\Theta}\bm{H}_{JR}\bm{w}_J^\star\|^2 \\& + \sigma_R^2\|\bm{\Theta}\|_F^2 \leq P_R^{\max},\qquad \end{aligned}  \\
    & \begin{aligned} P_J \geq  &\frac{\gamma|\bm{w}_D^\dagger\tilde{\bm{H}}_{SD}\bm{w}_{S}|}
        {\sqrt{C_J}\|\bm{w}_D^\dagger\tilde{\bm{H}}_{JD}\|} \\
    &-\frac{\sigma_R^2\|\bm{w}_D^\dagger\bm{H}_{RD}\bm{\Theta}\|^2+ \sigma_D^2}{\|\bm{w}_D^\dagger\tilde{\bm{H}}_{JD}\|^2} .\end{aligned}
\end{IEEEeqnarray}
We can easily verify that for the problem in~(\ref{eq:subprob-pwr}), it maximizes a concave objective function over a convex feasible set defined by these constraints, the transformed problem is a convex optimization problem. Therefore, it can be conveniently and efficiently solved by using standard off-the-shelf convex optimization tools, such as \texttt{CVX}. Once the optimal $\gamma^\star$ is found, the optimal legitimate transmit power is readily obtained as $P_S^\star = (\gamma^\star)^2$.

\subsection{Beamforming Optimization}

This subproblem focuses on optimizing the legitimate system's transmit and receive beamforming vectors, while fixing the other optimization variables. These vectors are constrained to be unit-norm, i.e., $\|\bm{w}_S\|^2=1$ and $\|\bm{w}_D\|^2=1$, and dictate the spatial direction of the legitimate transmission and reception. Note that the jammer's transmit beamforming vector is not explicitly optimized here as it is determined by the jammer's best response strategy in a closed form as discussed before.

Technically, the legitimate-side beamformers should be optimized to maximize the legitimate user's utility function, while strictly satisfying their unit-norm constraints and ensuring that the overall system constraints. However, jointly optimizing $\bm{w}_S$ and $\bm{w}_D$ while fully incorporating their complex coupling with these power constraints in~(\ref{eq:all-pwr-r}) and~(\ref{eq:all-pwr-j}) is a highly intricate non-convex problem. Here, we adopt a common simplification strategy. Particularly, we leverage two key observations: first, the beamformers are unit-norm and primarily affect the direction of signals, while their power level are determined by other variables. Second, the power constraints~(\ref{eq:all-pwr-r}) and~(\ref{eq:all-pwr-j}) have already undergone a tightening process in the problem reformulation, i.e., from~(\ref{eq:prob-l-alt}) to~(\ref{eq:prob-l-all}), which allows certain margins. Consequently, we strategically relax the explicit dependence of $\bm{w}_S$ and $\bm{w}_D$ on the power constraints in~(\ref{eq:all-pwr-r}) and~(\ref{eq:all-pwr-j}) during this specific beamforming optimization step. Instead, we focus on determining the optimal beamforming directions that maximize the SINR term in the objective function~(\ref{eq:all-obj}). This decoupling allows for well-known, low-complexity solutions for the beamformers. The rationale is that the iterative nature of the BCD algorithm will progressively refine all variables, and the satisfaction of power constraints is primarily enforced during the power and reflection optimizations.

In this regard, the transmission beamforming vector $\bm{w}_S$ is obtained based on the maximum ratio transmission principle, which maximizes the desired signal power by aligning $\bm{w}_S$ with the conjugate of the composite channel as
\begin{equation} \label{eq:opt-bf-s}
  \bm{w}_S^\star = \frac{\tilde{\bm{H}}_{SD}^\dagger\bm{w}_D}
            {\|\tilde{\bm{H}}_{SD}^\dagger\bm{w}_D\|}.
\end{equation}
Following the same principle of maximizing the SINR, for a fixed $\bm{w}_S$, the optimal receive beamforming vector $\bm{w}_D$ that maximizes the objective function can be found by solving a generalized Rayleigh quotient problem. This aims to enhance the desired signal component from $\tilde{\bm{H}}_{SD}\bm{w}_S$ while suppressing interference from $\tilde{\bm{H}}_{JD}\bm{w}_J^\star$ and receiver noise. The solution is given by the principal eigenvector of an equivalent matrix, given as
\begin{equation} \label{eq:opt-bf-d}
\bm{w}_D^\star = \frac{(\tilde{\bm{H}}_{JD}\tilde{\bm{H}}_{JD}^\dagger)^{-1}
        \tilde{\bm{H}}_{SD}\bm{w}_{S}}
            {\|(\tilde{\bm{H}}_{JD}\tilde{\bm{H}}_{JD}^\dagger)^{-1}
        \tilde{\bm{H}}_{SD}\bm{w}_{S}\|}.
\end{equation}

\subsection{Reflection Optimization}

Finally, we address the active RIS reflection optimization within the BCD framework, while all other system parameters are held constant from the previous BCD iterations. Accordingly, the reflection optimization subproblem, aiming to maximize the legitimate user's SINR-related objective function, is specified as
\begin{IEEEeqnarray} {cl} \label{eq:subprob-refl}
    \IEEEyesnumber \IEEEyessubnumber*
    \max_{\bm{\Theta}} 
        \quad&\frac{|\bm{w}_D^\dagger\tilde{\bm{H}}_{SD}\bm{w}_{S}|}
        {\|\bm{w}_D^\dagger\tilde{\bm{H}}_{JD}\|} \label{eq:subprob-refl-obj} \\
        \mathrm{s.t.}\quad&(\ref{eq:all-pwr-r}),~(\ref{eq:all-refl-amp}),~(\ref{eq:all-beam-j}),~(\ref{eq:all-pwr-j}), \nonumber
\end{IEEEeqnarray}
where the reflection coefficient matrix $\bm{\Theta}$ is embedded within the composite channel matrices. The constraints include the ARIS power budget, element-wise amplification limits, and the jammer's best response conditions. To simplify the problem, particularly the ARIS power constraint (\ref{eq:all-pwr-r}) which involves the jamming beamformer $\bm{w}_J^*$, we can leverage the fact that $\bm{w}_J^*$ is unit-norm, and apply the inequality $\|\bm{\Theta}\bm{H}_{JR}\bm{w}_J\|^2 \leq \|\bm{\Theta}\bm{H}_{JR}\|_F^2\|\bm{w}_J\|^2 = \|\bm{\Theta}\bm{H}_{JR}\|_F^2$. This step allows us to replace the term involving $\bm{w}_J^*$ with its upper bound related to the Frobenius norm, leading to a more tractable and tightened power constraint at the ARIS. This results in the following optimization problem
\begin{IEEEeqnarray} {cl} \label{eq:subprob-refl-lb}
    \IEEEyesnumber \IEEEyessubnumber*
    \max_{\bm{\Theta}} 
        \quad&\frac{|\bm{w}_D^\dagger\tilde{\bm{H}}_{SD}\bm{w}_{S}|}
        {\|\bm{w}_D^\dagger\tilde{\bm{H}}_{JD}\|}  \\
        \mathrm{s.t.}\quad &(\ref{eq:all-refl-amp}),~(\ref{eq:all-pwr-j}), \nonumber \\
        & \begin{aligned}  P_S\|\bm{\Theta}\bm{H}_{SR}\bm{w}_S\|^2  &+ P_J\|\bm{\Theta}\bm{H}_{JR}\|_F^2 \\& + \sigma_R^2\|\bm{\Theta}\|_F^2 \leq P_R^{\max}, \end{aligned} \label{eq:all-pwr-r-lb}
\end{IEEEeqnarray}
which provides a lower-bound solution to the original subproblem in~(\ref{eq:subprob-refl}) by relaxing the intricate handling of the unit-norm jamming beamformer within the power constraint.

To facilitate the subsequent analysis, we reorganize the notations such that the reflection matrix can be represented as a vector to reduce the dimensions. As such, for the objective function, we rewrite 
\begin{equation}
\begin{aligned}
    \bm{w}_D^\dagger\tilde{\bm{H}}_{SD}\bm{w}_{S}
    = \:& \bm{w}_D^\dagger(\bm{H}_{RD}\bm{\Theta}\bm{H}_{SR}+\bm{H}_{SD})\bm{w}_{S} \\
    = \:& \bm{w}_D^\dagger\bm{H}_{RD}\mathsf{diag}(\bm{H}_{SR}\bm{w}_{S})\bm{\theta}
    +\bm{w}_D^\dagger\bm{H}_{SD}\bm{w}_{S} \\
    = \:&  \bm{a}^\dagger\bm{\theta} + b,
\end{aligned}
\end{equation}
where
\begin{equation}
\bm{a}^\dagger \triangleq \bm{w}_D^\dagger\bm{H}_{RD}\mathsf{diag}(\bm{H}_{SR}\bm{w}_{S}),\quad b \triangleq \bm{w}_D^\dagger\bm{H}_{SD}\bm{w}_{S}.
\end{equation}
Similarly, we have
\begin{equation}
\begin{aligned}
    \bm{w}_D^\dagger\tilde{\bm{H}}_{JD}
    = \:&\bm{w}_D^\dagger(\bm{H}_{RD}\bm{\Theta}\bm{H}_{JR}+\bm{H}_{JD}) \\
    = \:& \bm{\theta}^T \mathsf{diag}(\bm{w}_D^\dagger\bm{H}_{RD})\bm{H}_{JR}
    +\bm{w}_D^\dagger\bm{H}_{JD} \\
    = \:&\bm{\theta}^T\bm{C} + \bm{d}^\dagger,
\end{aligned}
\end{equation}
where
\begin{equation}
  \bm{C} = \mathsf{diag}(\bm{w}_D^\dagger\bm{H}_{RD})\bm{H}_{JR},\quad  \bm{d}^\dagger = \bm{w}_D^\dagger\bm{H}_{JD}.
\end{equation}
Also, for the terms involving the reflection coefficient matrix within the constraints, we perform similar vectorizations as
\begin{equation}
   \bm{w}_D^\dagger\bm{H}_{RD}\bm{\Theta}
    = \bm{\theta}^\dagger \mathsf{diag}(\bm{w}_D^\dagger\bm{H}_{RD})
    =  \bm{\theta}^\dagger\bm{D}
\end{equation} 
with $\bm{D} \triangleq \mathsf{diag}(\bm{w}_D^\dagger\bm{H}_{RD})$, and similarly,
\begin{equation}
    \bm{\Theta}\bm{H}_{SR}\bm{w}_S 
    = \mathsf{diag}(\bm{H}_{SR}\bm{w}_S)\bm{\theta}, \quad
    \bm{\Theta}\bm{H}_{JR} 
    = \mathsf{diag}(\bm{\theta})\bm{H}_{JR}.
\end{equation}
Thenceforth, the optimization in terms of reflection vector is obtained as
\begin{IEEEeqnarray} {cl} \label{eq:subprob-refl-vec}
    \IEEEyesnumber \IEEEyessubnumber*
\max_{\bm{\theta}} 
    \quad&\frac{|\bm{a}^\dagger\bm{\theta} + b|}
    {\|\bm{\theta}^T\bm{C} + \bm{d}^\dagger\|}  \\
    \mathrm{s.t.}\quad & \begin{aligned} P_S\|\mathsf{diag}(\bm{H}_{SR}\bm{w}_S)\bm{\theta}\|^2 
    &+ P_J\|\mathsf{diag}(\bm{\theta})\bm{H}_{JR}\|_F^2 \\
    &+ \sigma_R^2\|\bm{\theta}\|^2 \leq P_R^{\max}, \qquad\qquad  \end{aligned} \label{eq:vec-pwr-r} \\
    &|\theta_n| \leq \alpha^{\max}, \quad \forall n\in \mathcal{N},  \label{eq:vec-amp} \\
    &P_J \geq  \frac{\sqrt{P_S}|\bm{a}^\dagger\bm{\theta} + b|}
        {\sqrt{C_J}\|\bm{\theta}^T\bm{C} + \bm{d}^\dagger\|}
    -\frac{\sigma_R^2\|\bm{\theta}^\dagger\bm{D}\|^2+ \sigma_D^2}
        {\|\bm{\theta}^T\bm{C} + \bm{d}^\dagger\|^2}.  \label{eq:vec-pwr-j}
\end{IEEEeqnarray}
For the problem in~(\ref{eq:subprob-refl-vec}), we can readily see that its non-convex nature presents a significant challenge. The non-convexity arises from multiple sources: the fractional objective function involving magnitudes of complex terms, the quadratic terms involving the reflection in constraint~(\ref{eq:vec-pwr-r}), and particularly the highly complex and non-convex constraint in~(\ref{eq:vec-pwr-j}) which represents the jammer's power reaction and reflections.

To address the difficulties in solving the problem, we mainly employ the successive convex approximation (SCA) technique, which involves iteratively solving a sequence of approximated convex problems. First, to handle the fractional objective function, we introduce the auxiliary variables
\begin{align} 
     \phi^2 &\leq  |\bm{a}^\dagger\bm{\theta} + b|, \label{eq:phi} \\
     \psi &\geq  \|\bm{\theta}^T\bm{C} + \bm{d}^\dagger\|,\label{eq:psi}
\end{align}
such that the objective function can be replaced by its lower bound as
\begin{equation}
    \max_{\bm{\theta},\phi,\psi} \quad \frac{\phi^2}{\psi},
\end{equation}
which is jointly convex with respect to $ \phi $ and $ \psi $. Then, we can apply the SCA technique at the point $ \left\{\phi_0, \psi_0\right\} $ and thus obtain a linear lower-bound counterpart as the objective function
\begin{equation}
\frac{\phi^2}{\psi} \geq \frac{2\phi_0\phi\psi_0-\phi_0^2\psi}{\psi_0^2}.
\end{equation}
Similarly, for the non-convexity in~(\ref{eq:phi}), we consider its equivalence for better tractability as
\begin{equation}
    \phi^4 \leq  |\bm{a}^\dagger\bm{\theta} + b|^2 
    = \bm{a}^\dagger\bm{\theta}\bm{\theta}^\dagger\bm{a}
    + 2\Re\{\bm{a}^\dagger\bm{\theta}b\}+|b|^2,
\end{equation}
where the convex right-hand side is replaced by the linear approximation at $ \bm{\theta}_0 $ such that
\begin{equation} \label{eq:phi-app}
    \phi^4 \leq  \bm{a}^\dagger(\bm{\theta}\bm{\theta}_0^\dagger
    +\bm{\theta}_0\bm{\theta}^\dagger-\bm{\theta}_0\bm{\theta}_0^\dagger)\bm{a}
    + 2\Re\{\bm{a}^\dagger\bm{\theta}b\}+|b|^2.
\end{equation}

Moreover, for the non-convex constraint in~(\ref{eq:vec-pwr-j}), we introduce several auxiliary variables to break it down into more manageable parts
\begin{align}
    \mu^2 &\geq |\bm{a}^\dagger\bm{\theta} + b|, \label{eq:mu}\\
    \nu &\leq \|\bm{\theta}^T\bm{C} + \bm{d}^\dagger\|, \label{eq:nu} \\
    \xi^2 &\leq \sigma_R^2\|\bm{\theta}^\dagger\bm{D}\|^2+ \sigma_D^2, \label{eq:xi} \\
    \zeta &\geq \|\bm{\theta}^T\bm{C} + \bm{d}^\dagger\|^2, \label{eq:zeta}
\end{align}
such that the constraint in~(\ref{eq:vec-pwr-j}) is tightened as
\begin{equation} \label{eq:pwr-j-alt}
    P_J \geq \frac{\sqrt {P_S}}{\sqrt{C_J}} \frac{\mu^2}{\nu}- \frac{\xi^2}{\zeta}.
\end{equation}
Each of the new constraints (\ref{eq:mu})-(\ref{eq:zeta}) and the transformed constraint (\ref{eq:pwr-j-alt}) might still be non-convex. To tackle the non-convexity in~(\ref{eq:mu}), we adopt the SCA technique such that the relationship involving $|\bm{a}^\dagger\bm{\theta} + b|$ is replaced by a convex surrogate. The constraint becomes
\begin{equation} \label{eq:mu-app}
    2\mu_0\mu-\mu_0^2 \geq |\bm{a}^\dagger\bm{\theta} + b|
\end{equation}
which is convex as approximated at $ \mu_0 $. Also, we have for the non-convex constraint in~(\ref{eq:nu})
\begin{equation} \label{eq:nu-ext}
\begin{aligned} 
    \nu^2 &\leq \|\bm{\theta}^T\bm{C} + \bm{d}^\dagger\|^2
    = (\bm{\theta}^T\bm{C}+\bm{d}^\dagger)(\bm{C}^\dagger\bm{\theta}^*+\bm{d}) \\
    &= \bm{\theta}^T\bm{C}\bm{C}^\dagger\bm{\theta}^*
            +2\Re\{\bm{\theta}^T\bm{C}\bm{d}\}+\|\bm{d}\|^2  \\
    &= \mathsf{Tr}(\bm{C}\bm{C}^\dagger\bm{\theta}^*\bm{\theta}^T)+2\Re\{\bm{\theta}^T\bm{C}\bm{d}\}+\|\bm{d}\|^2,
\end{aligned}
\end{equation}
where the right-hand side is approximated at $ \bm{\theta}_0 $ and thus the convexfied version of~(\ref{eq:nu-ext}) is obtained as
\begin{equation} \label{eq:nu-app}
    \begin{aligned}
        \nu^2 \leq \mathsf{Tr}(\bm{C}\bm{C}^\dagger(\bm{\theta}^*\bm{\theta}_0^T
            +\bm{\theta}_0^*\bm{\theta}^T-\bm{\theta}_0^*\bm{\theta}_0^T)) \\
            +2\Re\{\bm{\theta}^T\bm{C}\bm{d}\}+\|\bm{d}\|^2.
    \end{aligned}   
\end{equation}
Similarly, for the non-convex constraint in~(\ref{eq:xi}), we have
\begin{equation}
    \begin{aligned}
    \xi^2 \leq  \sigma_R^2\|\bm{\theta}^\dagger\bm{D}\|^2+ \sigma_D^2  
    &= \sigma_R^2(\bm{\theta}^\dagger\bm{D}\bm{D}^\dagger\bm{\theta})+ \sigma_D^2  \\
    &=  \sigma_R^2\mathsf{Tr}(\bm{D}\bm{D}^\dagger\bm{\theta}\bm{\theta}^\dagger)+ \sigma_D^2,
\end{aligned}
\end{equation}
which is convexified by the first-order approximation as
\begin{equation} \label{eq:xi-app}
    \xi^2 \leq \sigma_R^2\mathsf{Tr}(\bm{D}\bm{D}^\dagger(\bm{\theta}\bm{\theta}_0^\dagger
    +\bm{\theta}_0\bm{\theta}^\dagger-\bm{\theta}_0\bm{\theta}_0^\dagger))+ \sigma_D^2.
\end{equation}
Finally, for the constraint in~(\ref{eq:pwr-j-alt}), the inherent non-convexity is also tackled through SCA technique, leading to the convex counterpart as
\begin{equation} \label{eq:pwr-j-app}
     P_J \geq \frac{\sqrt {P_S}}{\sqrt{C_J}} \frac{\mu^2}{\nu}-\frac{2\xi_0\xi\zeta_0-\xi_0^2\zeta}{\zeta_0^2}.
\end{equation}

With the systematic application of auxiliary variables and SCA to all non-convex components of the objective function and constraints derived from~(\ref{eq:subprob-refl-vec}), we arrive at an lower-bounded and tractable convex optimization problem as
\begin{IEEEeqnarray} {cl} \label{eq:subprob-refl-vec-app}
    \IEEEyesnumber \IEEEyessubnumber*
\max_{\bm{\theta},\phi,\psi,\mu,\nu,\xi,\zeta} 
        &\quad \frac{2\phi_0\phi\psi_0-\phi_0^2\psi}{\psi_0^2} \\
        \mathrm{s.t.} &\quad (\ref{eq:vec-pwr-r}),~(\ref{eq:vec-amp}), \nonumber\\
        &\quad (\ref{eq:psi}),~(\ref{eq:phi-app}), \nonumber\\
        &\quad (\ref{eq:zeta}),~(\ref{eq:mu-app}),~(\ref{eq:nu-app}),~(\ref{eq:xi-app}), \nonumber\\
        &\quad (\ref{eq:pwr-j-app}),\nonumber
\end{IEEEeqnarray}
which is approximated at $ \left\{ \bm{\theta}_0, \phi_0, \psi_0, \mu_0, \xi_0, \zeta_0 \right\} $. Then, we solve a series of convex problems in the form of~(\ref{eq:subprob-refl-vec-app}), and update the approximation point by the obtained optimum. The obtained sequence through SCA is guaranteed to converge to a Karush-Kuhn-Tucker (KKT) point, providing a locally optimal solution to the original non-convex reflection optimization subproblem in~(\ref{eq:subprob-refl-vec}), or equivalently in~(\ref{eq:subprob-refl-lb}).

\subsection{Overall Algorithm and Equilibrium Workflow}

In preceding discussions, we have elaborated the solution to the subproblems to determine the anti-jamming strategy, including the power allocation, beamforming, and reflection. Then, the solution to the subproblems are iterated in the BCD framework until the convergence. The overall algorithm is for deriving the legitimate-side's strategy is summarized in Alg.~\ref{alg}.

\begin{algorithm}[t] 
\SetKwComment{Comment}{$\triangleright$\ }{}
\KwIn{Channel matrices, noise variances, maximum transmit and jamming powers, cost coefficients, ARIS power and amplification threshold;}
\KwOut{Legitimate-side strategy $\{P_S^*, \bm{w}_S^\star,\bm{w} _D^\star, \bm{\Theta}^\star\}$.}
\BlankLine
Initialize $P_S^{(0)}, \bm{w}_S^{(0)}, \bm{w}_D^{(0)}, $ satisfying the constraints\;
Calculate the jammer's anticipated strategy $P_J^{(0)}$ and $ \bm{w}_J^{(0)} $ using (\ref{eq:opt-j-bf}) and~(\ref{eq:opt-j-pwr})\;
\Repeat{$|u_L^{(0)} - u_L^{\star}| < \epsilon$ }{
  Calculate initial legitimate utility $u_L^{(0)}$ \;
  \Comment{Power Optimization}
  Solve the problem in~(\ref{eq:subprob-pwr}), and get $ P_J^{\star}$ and $ \gamma^{\star} $, then obtain $ P_S^\star \gets (\gamma^\star)^2 $\;
  $ P_S^{(0)} \gets P_S^{\star} $; $ P_J^{(0)} \gets P_J^{\star} $\;
  \Comment{Beamforming Optimization}
  Given $P_S^{(0)}$, $P_J^{(0)}$, and $\bm{\Theta}^{(0)}$, update $ \bm{w}_J $, $ \bm{w}_S $, $ \bm{w}_D $ by using (\ref{eq:opt-j-bf}), (\ref{eq:opt-bf-s}), and~(\ref{eq:opt-bf-d}), respectively\;
  $ \bm{w}_J^{(0)} \gets \bm{w}_J^{\star} $, $ \bm{w}_S^{(0)} \gets \bm{w}_S^{\star} $, $ \bm{w}_D^{(0)} \gets \bm{w}_D^{\star} $\;
  \Comment{Reflection Optimization}
  \Repeat{$ \| \eta^{(0)} - \eta^\star \| \le \varepsilon $ }{
  Calculate the objective function in~(\ref{eq:subprob-refl-obj}) as $ \eta^{(0)} $ by using $ \bm{w}_S^{(0)}$, $ \bm{w}_D^{(0)}$, and $\bm{\Theta}^{(0)}$\;
  Solve the problem in~(\ref{eq:subprob-refl-vec-app}) at the point $ \left\{ \bm{\theta}_0, \phi_0, \psi_0, \mu_0, \xi_0, \zeta_0 \right\} $ and obtain the updated objective as $ \eta^\star $ along with the solution $ \bm{\theta}^\star,\phi^\star,\psi^\star,\mu^\star,\nu^\star,\xi^\star,\zeta^\star $\;
  $\bm{\theta}_0 \gets \bm{\theta}^\star$, $\phi_0 \gets \phi^\star$, $\psi_0 \gets \psi_\star$, $\mu_0 \gets \mu^\star$, $\xi_0 \gets \xi^{\star}$, $\zeta_0 \gets \zeta^\star$\;
  }
  Reconstruct $ \bm{\Theta}^{\star} $ from $ \bm{\theta}^\star $, and $ \bm{\Theta}^{(0)} \gets \bm{\Theta}^\star $\;
  \Comment{Utility Update and Convergence}
  Recalculate the jammer's strategy and calculate the legitimate utility as $ u_L^\star $\;
}
\caption{Algorithm for Legitimate-Side Strategy}
\label{alg}
\end{algorithm}

Therefore, we have derived the optimal strategies for the legitimate side and the jammer in the game, which correspond to the conditions in~(\ref{eq:eq-j}) and~(\ref{eq:eq-l}). Then, the individual optimality guides the game players to achieve the Stackelberg equilibrium. Specifically, For the legitimate system, as the leader, initiates the decision-making process. It intelligently anticipates the optimal reaction of the jammer as the follower. In this regard, the anticipated follower behavior is embedded as a constraint or incorporated into the objective function of the leader's optimization problem. The leader then solves this problem, by using Alg.~\ref{alg}, to determine its optimal transmission strategy. This strategy is inherently robust against the jammer's best possible counterattack. Once the legitimate system commits to and implements its strategy, the jammer observes these actions. The jammer then employs its optimal strategy by selecting its jamming beamformer and power. This response maximizes the jammer's utility given the leader's actions. Finally, the combined set of strategies of both players constitutes the Stackelberg equilibrium. At this equilibrium, neither the leader nor the follower has an incentive to unilaterally deviate from their chosen strategies. The leader has already optimized its outcome knowing the follower will react optimally, and the follower is indeed playing its best response to the leader's actions. The jammer's realized strategy aligns with the anticipated reaction that the leader factored into its own strategic calculations, thereby fulfilling the equilibrium conditions specified in~(\ref{eq:eq-j}) and~(\ref{eq:eq-l}).

The proposed algorithm is guaranteed to converge to a stationary point of the leader's problem (and thus the equilibrium of the game). This is because in each BCD iteration, the subproblems (power, beamforming, reflection) are solved to the (local) optimality, and the leader's utility function is thus non-decreasing after each update within a compact feasible set. Regarding the complexity, the BCD iterations incorporate the power, beamforming, and reflection optimization. Specifically, the power optimization concerns fixed two variables, and thus the cost is negligible. The beamforming is determined with closed-form solutions, where the transmit beamforming is in maximum-ratio manner and has a complexity of $\mathcal{O}(N_S N_D)$, and the receive beamforming has matrix inverse operations and thus the complexity is $\mathcal{O}\left(N_D^3\right)$. For the reflection optimization, the SCA procedure is adopted, where each update follows interior-point methods and has a complexity of $\mathcal{O}\left(N^{3.5}\right)$. Finally, denote the numbers of iterations required for the outer BCD and inner SCA loops as $I_{BCD}$ and $I_{SCA}$, respectively, the overall complexity is $\mathcal{O}(I_{BCD} (N_S N_D + N_D^3 + (I_{SCA} N^{3.5}))$.

\subsection{Implementation Issue}
While the proposed algorithm achieves the game equilibrium, the practical algorithm deployment brings along several implementation issues detailed below. First, we assume perfect channel state information in this work, which requires the acquisition of high-dimensional cascaded channels related to both players. This requires channel training among the transceivers and further information feedback. Second, the system requires stringent time and phase synchronization to ensure coherent signal combination, and thus the RIS configuration needs to be finished within the coherent time. Third, we assume ideal linear amplifiers, the implementation needs to specify the (near-)linear power interval and make sure the RIS amplification stays in the proper region.

\section{Simulation Results} \label{sec:sim}

In this section, we present the simulation results to demonstrate the performance and advantages of the proposed scheme. For the simulation, we consider that the legitimate source, destination, and jammer are located at the coordinates (0, 100), (400, 0), and (100, 400), respectively (distances in meters). The active RIS is located at (200, 0) with a height of 200. The legitimate source node and the jammer have 4 transmit antennas, and the legitimate destination node has 2 receive antennas. The active RIS has 50 reflecting elements. For each element, the maximum amplitude is 10~dB, with an overall power budget of 20~dBm. The maximum legitimate transmit power and jamming power are 5~W and 10~W, respectively. The ground channel gain suffers a path-loss exponent of 3.5 with Rayleigh fading, and channels related to the RIS follow a Rician model with a large-scale path-loss exponent of 2.3 and a Rician factor of 10~dB. The background noise at all nodes is -140~dBW. The power cost coefficients at the legitimate source node and the jammer are 5 and 6, respectively.

\subsection{Equilibrium Status Evaluation}

\begin{figure}[t]
    \centering
    \includegraphics[width=0.9\linewidth]{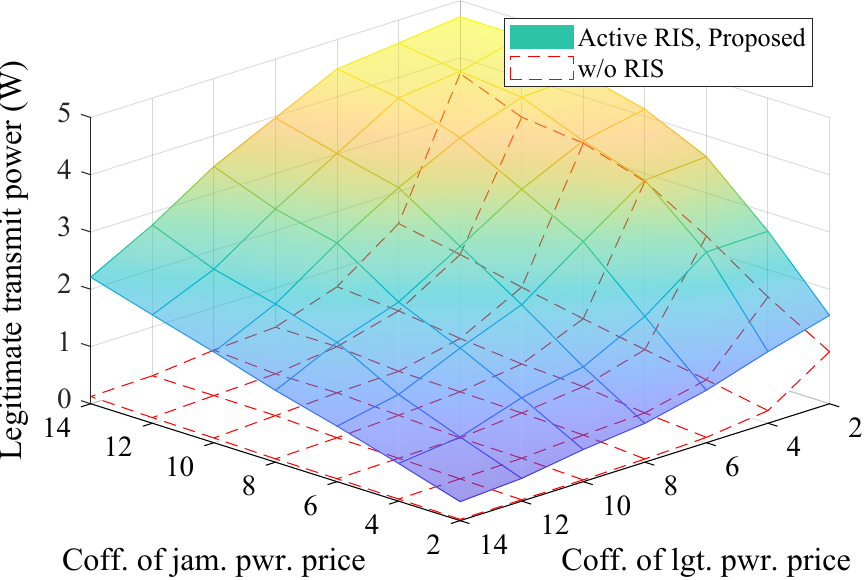}
    \caption{Legitimate transmit power w.r.t. power cost coefficients.}
    \label{fig:ps_costs}
\end{figure}

\begin{figure}[t]
    \centering
    \includegraphics[width=0.9\linewidth]{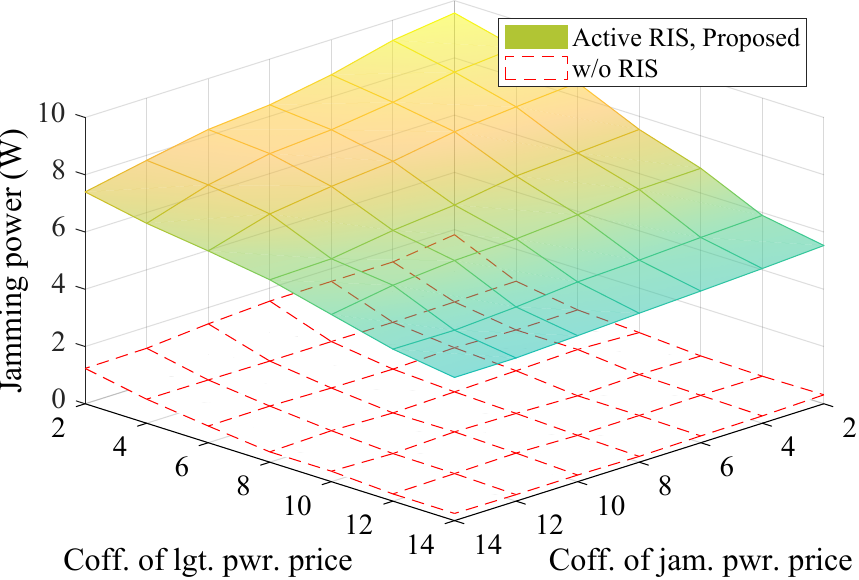}
    \caption{Jamming power w.r.t. power cost coefficients.}
    \label{fig:pj_costs}
\end{figure}

Figs.~\ref{fig:ps_costs} and~\ref{fig:pj_costs} illustrate the equilibrium legitimate transmit power and jamming power, respectively, as functions of their power cost coefficients. In Fig.~\ref{fig:ps_costs}, we observe that for both the proposed scheme and the baseline scheme without RIS, the legitimate transmit power decreases as its own power price increases, which is an intuitive outcome reflecting the higher cost of transmission. Conversely, when the jammer's power price increases, the legitimate user tends to increase its transmit power. This behavior highlights the strategic advantage of the legitimate user as the Stackelberg leader, as jamming becomes more costly for the adversary, the leader exploit the opportunity by boosting its own signal strength. The superiority of the proposed scheme is clearly demonstrated, as it consistently enables the legitimate user to operate at significantly higher transmit power levels across the entire range different power costs compared to the scenario without RIS. This indicates that the active RIS effectively enhances the anti-jamming communication link, making higher power transmission more beneficial and viable for the legitimate system.

\begin{figure}[t]
    \centering
    \includegraphics[width=0.9\linewidth]{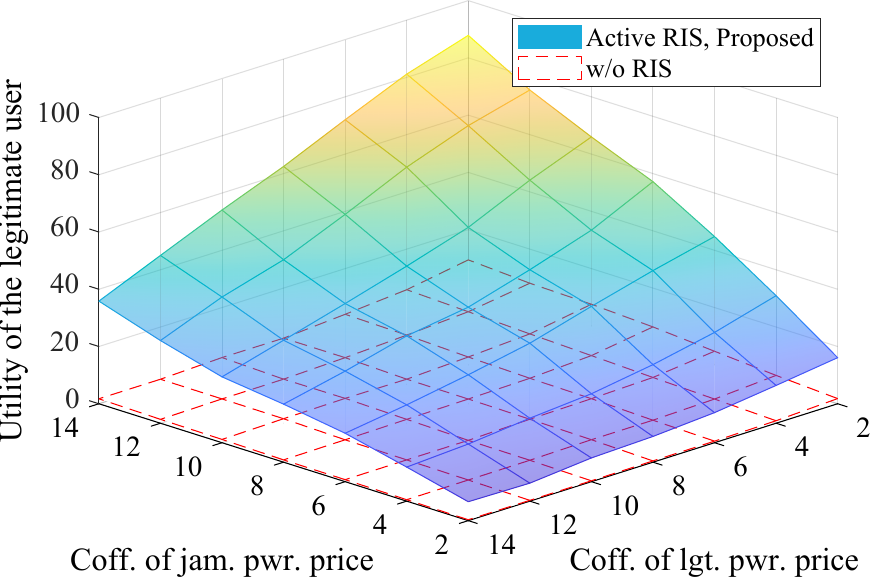}
    \caption{Legitimate utility w.r.t. power cost coefficients.}
    \label{fig:ul_costs}
\end{figure}

\begin{figure}[t]
    \centering
    \includegraphics[width=0.9\linewidth]{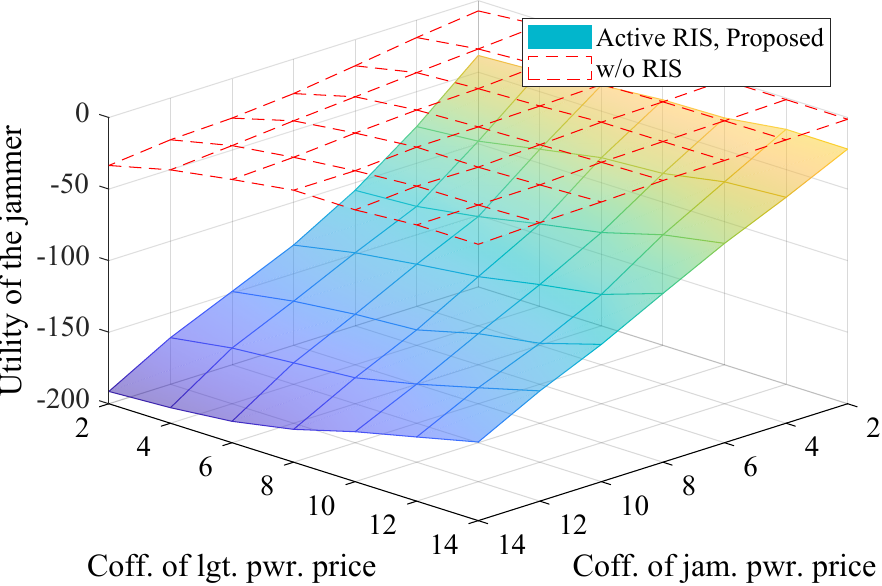}
    \caption{Jamming utility w.r.t. power cost coefficients.}
    \label{fig:uj_costs}
\end{figure}

Fig.~\ref{fig:pj_costs} shows the jammer's power allocation as the follower's response. As anticipated, the jamming power decreases substantially when the jammer's own power price rises for both schemes. Interestingly, an increase in the legitimate user's power price also leads to a reduction in jamming power. This is because a higher legitimate power cost typically forces the legitimate user to lower its power, and the jammer as a follower, adapts its strategy to the less aggressive legitimate transmission. Comparing the two schemes, the proposed method often faces similar or even slightly reduced jamming power compared to the no-RIS case, particularly when jamming cost is not excessively high. This, coupled with the legitimate user's ability to maintain higher power as shown in Fig.~\ref{fig:ps_costs}, underscores the efficacy of the active RIS in not only improving the desired signal path but also in mitigating the impact of jamming, thereby potentially mitigating the jamming behavior and effects. The advantage of the proposed method lies in its ability to intelligently manipulate the radio environment via the reflection and amplification, allowing the legitimate user to maintain a superior communication posture and achieve a better utility.

In accordance with the transmit and jamming behavior, Figs.~\ref{fig:ul_costs} and \ref{fig:uj_costs} depict the utility of the legitimate user and the jammer, respectively, against the same power price coefficients. Fig.~\ref{fig:ul_costs} reveals that the legitimate user's utility consistently and significantly benefits from the proposed active RIS-assisted scheme, achieving substantially higher utility values compared to the no-RIS scenario across considered coefficient ranges. As the legitimate user's power price increases, the utility tends to decrease, which is expected since transmission becomes more costly, directly impacting the utility function. Conversely, when the jammer's power price increases, the utility generally improves. This correlates with the observations from Figs.~\ref{fig:ps_costs} and~\ref{fig:pj_costs}, that a higher jamming cost leads to reduced jamming power and allows the legitimate user to increase its power, thereby enhancing the SINR component of its utility while the jammer is less effective. The superior performance of the ARIS scheme in terms of legitimate utility reveals its ability to create a more favorable communication environment that induces better outcomes for the legitimate communication system.

Fig.~\ref{fig:uj_costs} illustrates the jammer's utility. In accordance with its definition, the jammer's utility is negatively impacted by the SINR achieved by the legitimate user and its own power consumption. Consequently, as the jammer's power price increases, its utility improves because it reduces its jamming power, as seen in Fig.~\ref{fig:pj_costs}, thus lowering its costs. When the legitimate user's power price increases, the jammer's utility also tends to improve. This is because a higher leads to a lower power by the legitimate user, resulting in a lower SINR, which benefits the jammer. Notably, the jammer consistently achieves a worse utility when the proposed active RIS is employed compared to the no-RIS case. This is a direct consequence of the active RIS enhancing the legitimate user's SINR, which is detrimental to the jammer's objective. This demonstrates the dual benefit of the proposed scheme, that it not only boosts the legitimate user's performance but also more effectively contains the jammer's ability to disrupt the communications, and thus more effectively protect the legitimate communications.

\subsection{Performance Comparison}

\begin{figure}[t]
    \centering
    \includegraphics[width=0.9\linewidth]{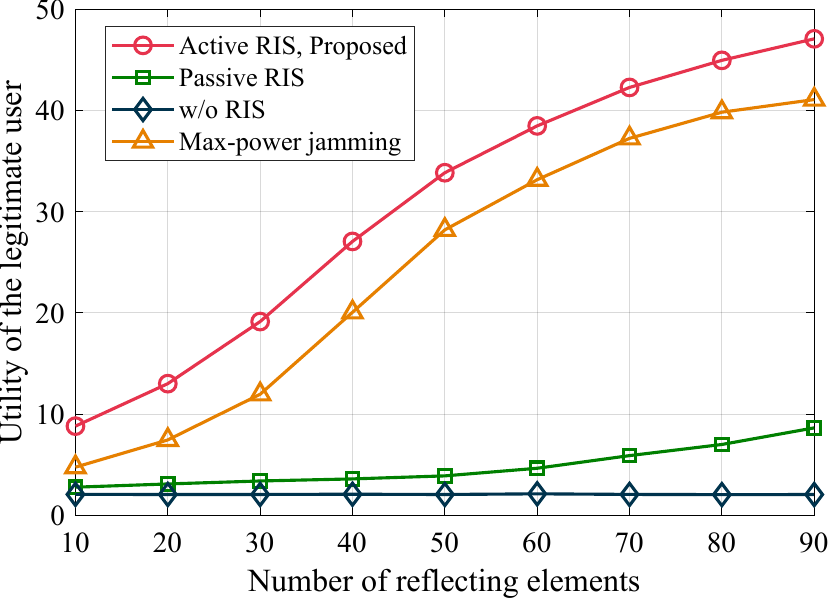}
    \caption{Legitimate utility w.r.t. number of reflecting elements.}
    \label{fig:ul_N}
\end{figure}

\begin{figure}[t]
    \centering
    \includegraphics[width=0.9\linewidth]{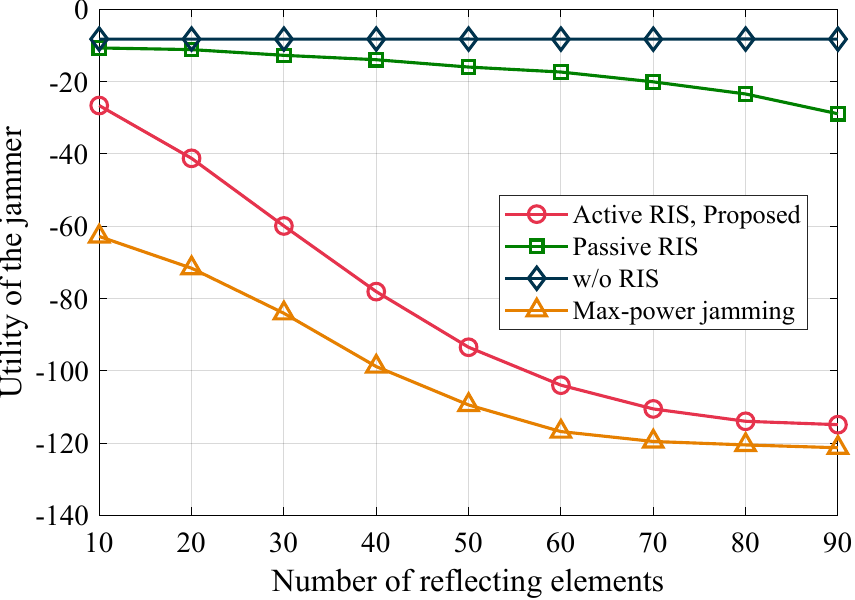}
    \caption{Jamming utility w.r.t. number of reflecting elements.}
    \label{fig:uj_N}
\end{figure}

\begin{figure}[t]
    \centering
    \includegraphics[width=0.9\linewidth]{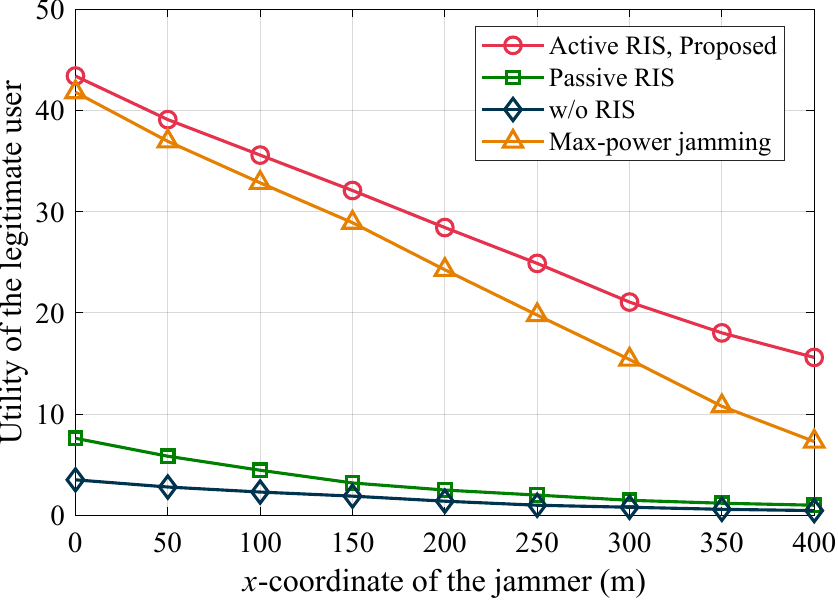}
    \caption{{Legitimate utility w.r.t. location of jammer.}}
    \label{fig:ul_xJ}
\end{figure}

\begin{figure}[t]
    \centering
    \includegraphics[width=0.9\linewidth]{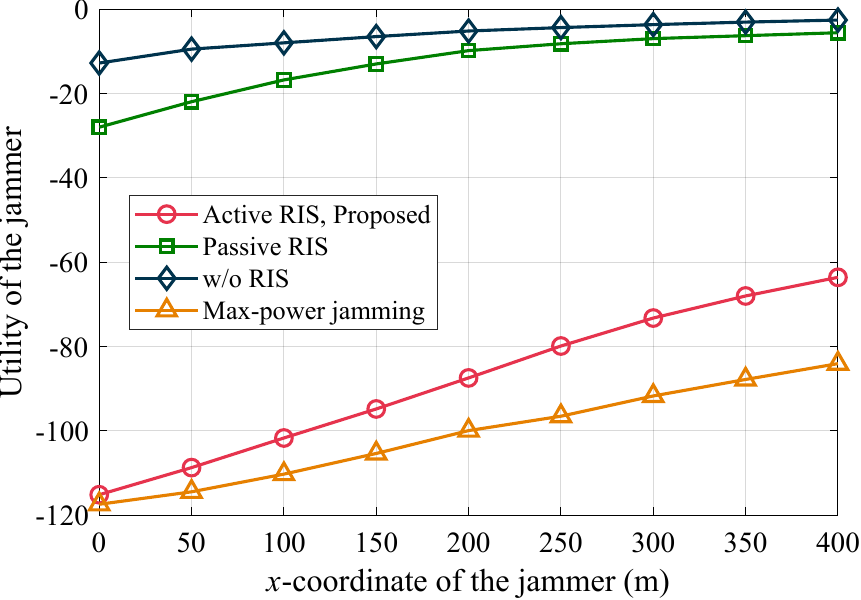}
    \caption{{Jamming utility w.r.t. location of jammer.}}
    \label{fig:uj_xJ}
\end{figure}

Figs.~\ref{fig:ul_N} and~\ref{fig:uj_N} demonstrate the impact of the number of reflecting elements on the utilities of the legitimate user and jammer, respectively. Here, we consider four schemes: 1) the proposed, 2) a conventional passive RIS without amplification, 3) a baseline scenario with no RIS, 4) the jammer always uses its maximum power and only adapts the beamforming to react to the legitimate transmissions.

We first show the performance considering different reflecting elements at the active RIS. As shown in Fig.~\ref{fig:ul_N}, the legitimate user's utility significantly benefits from an increasing number of reflecting elements for the cases with an active RIS. This is expected, as more elements provide higher degrees of freedom for manipulating the radio environment to enhance the desired signal and suppress jamming signals. The proposed scheme consistently outperforms all other schemes, achieving the highest utility. The substantial advantage over passive RIS, highlights the benefit of signal amplification at the RIS elements, which helps to overcome the "double fading" effect more effectively. Both RIS-assisted schemes yield considerably higher utility than the no-RIS case, emphasizing the fundamental advantage of employing RIS technology. For max-power jamming scenario, the legitimate side still can fully exploit the spatial diversity from tranceiving antennas and RIS adjustment to mitigate the jamming signal, and thus maintains relatively high utility.

Fig.~\ref{fig:uj_N} presents the jammer's utility. For the proposed active RIS scheme, the jammer's utility generally degrades as $N$ increases, for which the degradation is much more evident than the case with a passive RIS. This indicates that as the RIS becomes more effective with more elements, it becomes harder for the jammer to achieve its objectives. This validates the superior anti-jamming capabilities of proposed scheme by actively amplifying the desired signal and optimizing reflections, as it significantly enhances the legitimate user's SINR, and thus undermines the jamming effects. When without a RIS, the limited degree of freedom of the legitimate system allows the jammer to achieve a high jamming utility. For the case of max-power jamming, it induces the worst results in terms of jamming utility. This is because, as the active RIS significantly enhances the legitimate system's capability in intervening the radio environment, the maximum jamming power is not that effective, which is also shown in Fig.~\ref{fig:ul_N}. Meanwhile, the maximum jamming power induces a high cost, and thus the overall jamming utility becomes severely affected. These results collectively demonstrate the significant advantages of the proposed active RIS-assisted game strategy, not only in maximizing the legitimate system performance but also in effectively mitigating the jamming.

Figs.~\ref{fig:ul_xJ} and~\ref{fig:uj_xJ} evaluate the performance of the considered schemes as the jammer's location varies. As the jammer moves rightwards, it is located farther from the legitimate source while more closely approaching the legitimate destination. As observed in Fig.~\ref{fig:ul_xJ}, the legitimate user's utility generally decreases as the jammer moves closer to the destination. This is intuitive as a jammer positioned nearer to the receiver can induce a more significant jamming effect, making it harder for the legitimate system to achieve a high SINR. Across all jammer locations, the proposed active RIS scheme consistently provides the highest utility for the legitimate user. The performance gap between active RIS and passive RIS underscores the crucial role of signal amplification in combating jamming. Even when the jammer is very close to the destination, where it can be rather challenging to mitigate the attacks, the active RIS maintains a clear advantage, demonstrating its robustness. Notably, for the case of using maximum jamming power, we can similarly exploit the active RIS to alleviate the jamming effect, thus maintaining the advantages of the proposed scheme.

Also, Fig.~\ref{fig:uj_xJ} shows results for the jammer's utility. Generally, the jammer's utility tends to improve as it moves closer to the legitimate destination. This is because its jamming signals become more effective at degrading the legitimate user's SINR with shorter propagation paths to the destination. However, the proposed active RIS scheme consistently forces the jammer to achieve low utility, irrespective of its location. This highlights the effectiveness of our proposal in creating an unfavorable environment for the jammer. In contrast, the capability of a passive RIS to reduce the jammer's utility is rather limited, let alone the case without reflection and relying purely on antenna diversity. For the use of maximum jamming power, similar to preceding results, the high jamming power is not that effective in jamming attacks but induces a high power cost, and thus the achieved utility at the jammer induces the worst results. Thus, the ability of the proposed active RIS scheme to consistently yield high utility for the legitimate user while simultaneously minimizing the jammer's utility, regardless of the jammer's proximity to the source or destination, further validates its superiority and practical applicability in dynamic jamming environments.

\section{Conclusion} \label{sec:con}

In this paper, we investigated anti-jamming communications by employing an active RIS. We formulated the strategic interaction between the legitimate system and the jammer as a Stackelberg game. By confirming the existence of the equilibrium, we derived the equilibrium through backward induction, where a joint optimization of power allocation, transceiving beamforming, and active reflection was proposed to achieve the anti-jamming transmission strategy. Simulation results were presented to validate the effectiveness of the proposed anti-jamming design as compared with baselines under various communication scenarios. The results underscore the substantial benefits of incorporating active RIS technology and employing a strategic game approach to counteract intelligent jamming attacks. The ability of the active RIS to dynamically reshape the radio environment, coupled with the strategic foresight provided by the Stackelberg game model, manifests as an effective combination for enhancing wireless communication security and reliability.

\bibliographystyle{IEEEtran}
\bibliography{main}

\end{document}